\documentclass{PoS}
\usepackage{graphicx}

\title{Exploring the powering source of the TeV X-ray binary LS 5039}

\ShortTitle{Exploring the powering source of the TeV X-ray binary LS 5039}

\author{\speaker{Javier Mold\'on}, Marc Rib\'o, Josep M. Paredes\\
Departament d'Astronomia i Meteorologia and Institut de Ci\`encies del Cosmos (ICC), Universitat de Barcelona\\
Mart\'{\i} i Franqu\`es, 1\\
08028 Barcelona, Spain\\
E-mail: \email{jmoldon@am.ub.es}, \email{mribo@am.ub.es}, \email{jmparedes@ub.edu}}

\author{Josep Mart\'i\\
        Departamento de F\'isica (EPS), Universidad de Ja\'en\\
        E-mail: \email{jmarti@ujaen.es}}
        
\author{Maria Massi\\
        Max Planck Institut für Radioastronomie\\
        E-mail: \email{mmassi@mpifr-bonn.mpg.de}}

\abstract{LS 5039 is one of the four TeV emitting X-ray binaries detected up to
now. The powering source of its multi-wavelength emission can be accretion in a
microquasar scenario or wind interaction in a young non-accreting pulsar
scenario. These two scenarios predict different morphologic and peak position
changes along the orbital cycle of 3.9 days, which can be tested at
milliarcsecond scales using VLBI techniques. Here we present a campaign of
5~GHz VLBA observations conducted in June 2000 (2 runs five days apart). The
results show a core component with a constant flux density, and a fast change
in the morphology and the position angle of the elongated extended emission,
but maintaining a stable flux density. These results are difficult to fit
comfortably within a microquasar scenario, whereas they appear to be compatible
with the predicted behavior for a non-accreting pulsar.}

\FullConference{VII Microquasar Workshop: Microquasars and Beyond\\
		 September 1-5 2008\\
		 Foca, Izmir, Turkey}

\begin{document}

\section{Introduction} \label{introduction}

Very High Energy (VHE) gamma-ray emission in the TeV range has been detected in
four massive X-ray binaries. PSR~1259$-$63 contains a young non-accreting
millisecond (ms) radio pulsar \cite{aharonian05}. The accreting/ejecting
microquasar Cygnus~X-1 has recently been found to be a TeV emitter
\cite{albert07}. LS~I~+61~303 \cite{albert06}, suggested to be a fast
precessing microquasar \cite{massi04}, displays a changing milliarcsecond (mas)
radio morphology that Dhawan et al.\ \cite{dhawan06} interpreted in the context of the
interaction between the wind of the companion and the relativistic wind of a
young non-accreting ms pulsar \cite{dubus06}. Finally, the nature of the
powering source in LS~5039, whether accretion or rotation, is unknown.

\section{The binary system LS~5039} \label{system}

%-----------------------------------------------------------------------------
\begin{table}[b!] \caption[]{Orbital parameters of the binary system and
estimated stellar properties from \cite{casares05}.}
\label{tab:orbital} \begin{center} \begin{tabular}{cllr}
\hline\noalign{\smallskip} Symbol          &  & 
Parameter                                         &  value \\
\hline\noalign{\smallskip} $P_{\rm orb}$   &  &  Orbital period,
[days]                            &  $3.906\,03\pm0.000\,17$ \\ $T_{0}$        
&  &  Periastron passage, HJD$-$2400000                 &  $51943.09\pm0.10$ \\
$e$             &  &  Eccentricity                                      & 
$0.35\pm0.04$ \\ $\omega$        &  &  Argument of periastron,
[$^{\circ}$]              &  $225.8\pm3.3$ \\ $\gamma$        &  &  Systemic
radial velocity, [km~s$^{-1}$]           &  $17.2\pm0.7$ \\ $K_{1}$         & 
&  Velocity semi-amplitude  [km~s$^{-1}$]            &  $25.2\pm1.4$ \\
$a_{1}\sin(i)$  &  &  Projected semi-major axis, [${R}_{\odot}$]        & 
$1.82\pm0.10$ \\ $f(M)$          &  &  Mass function,
[${M}_{\odot}$]                    &  $0.0053\pm0.0009$ \\ $v~\sin(i)$     & 
&  Projected rotational velocity, [km~s$^{-1}$]& $113\pm8$ \\

  \noalign{\smallskip}\hline
$d$             &  & System distance, [kpc]                             & $2.5\pm0.5$ \\
$M_{\rm x}$     &  & Compact object mass, [${M}_{\odot}$]               & $1.49$--$1.81>M>8$--$10$ \\
$M_{\star}$     &  & Stellar mass, [${M}_{\odot}$]                      & $22.9_{-2.9}^{+3.4}$ \\
$R_{\star}$     &  & Stellar radius, [${R}_{\odot}$]                    & $9.3_{-0.6}^{+0.7}$ \\
$L_{\star}$     &  & Stellar luminosity, [erg~s$^{-1}$]                 & $7\times10^{38}$ \\
$\dot{m}_{\rm w}$& & Stellar mass loss rate, [${M}_{\odot}$~yr$^{-1}$]  & 7$\times10^{-7}$ \\
%$T_{\star}$     &  &  Star temperature, [K]                             & $39~000\pm1000$ \\
%Log~{\rm g}     &  &  log gravity                                       & $3.85\pm0.10$ \\
  \noalign{\smallskip}\hline
  \end{tabular}
  \end{center}

\end{table}
%-----------------------------------------------------------------------------

The optical star LS~5039 is a main-sequence young star with a luminosity
$L\simeq1.8\times10^{5}$~${L}_{\odot}$, and an effective temperature of $T_{\rm
eff}=39~000$~K. The visual magnitude of the star is $\sim11.3$. Approximate
values of the radius and mass are $R\simeq9$~${R}_{\odot}$\ and
$M\simeq23$~${M}_{\odot}$. The complete spectral type of the star is
ON6.5~V((f)). The orbital parameters of the system obtained from the
spectroscopic observations in \cite{casares05} are summarized in
Table~\ref{tab:orbital}. The mass of the compact object depends on the
inclination of the orbit, which can be between 11 and 75$^{\circ}$, and is in
the range 1.5--8~${M}_{\odot}$. The compact object can thus be a black hole or
a neutron star. Based on the mass function and inclination, the probability of
being a black hole is $\sim20$\%. The rotational velocity of the optical star
sets the lower limit of the inclination, and the absence of X-ray eclipses sets
the upper limit, assuming that the emission is produced in the vicinity of the
compact object. This upper limit can be higher if the emission takes place far
away from the compact object. In this slightly eccentric system, the objects
are separated 0.1~AU at periastron and 0.2~AU at apastron. In case the optical
companion is pseudo-synchronized, i.e.\ its rotational and orbital angular
velocities are synchronized at the periastron, the compact object mass would be
in the range 3.14--4.35~${M}_{\odot}$ (corresponding to an inclination of
$i=24\rlap.^{\rm\circ}9\pm2\rlap.^{\rm \circ}8$), and therefore it would be a
black hole. For a system like LS~5039, the synchronization timescale is
$\sim1$~Myr after the supernova explosion.

\begin{figure} \center \includegraphics[angle=0,scale=0.4]{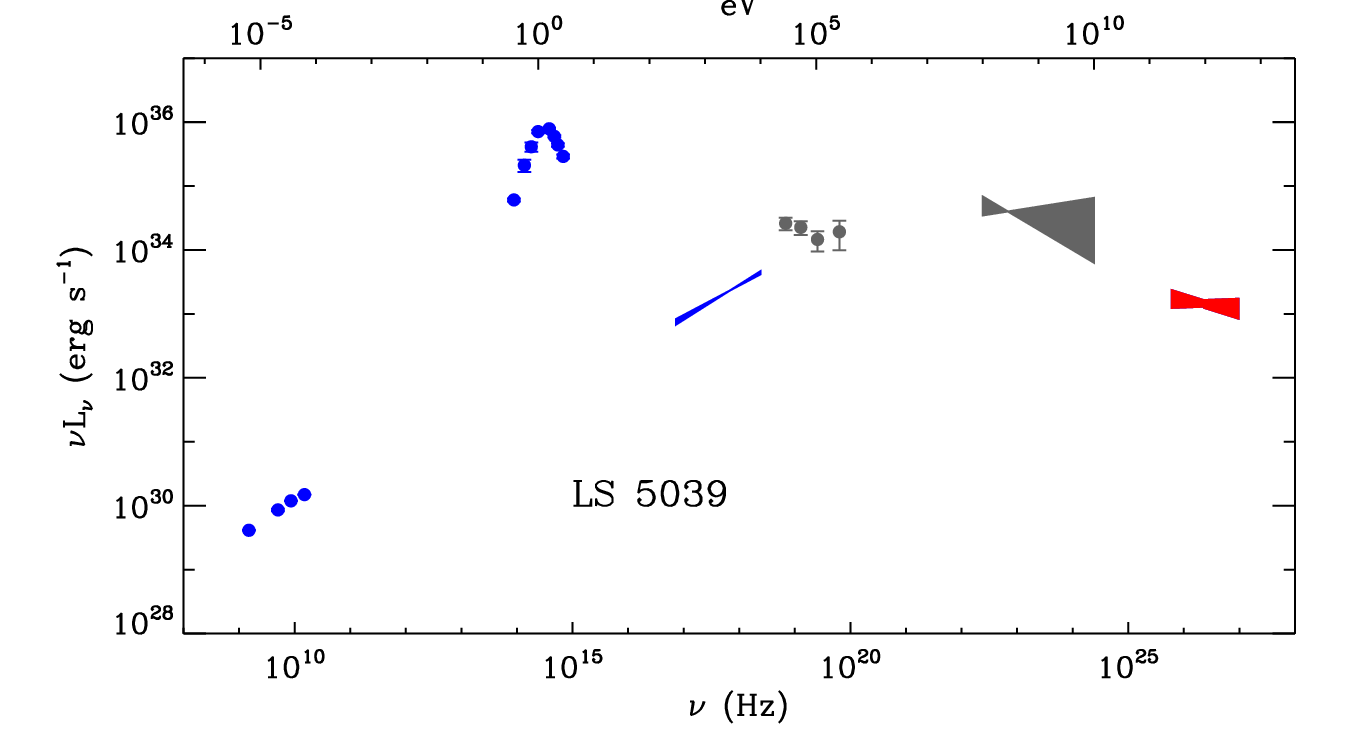}
\caption{Spectral energy distribution of LS~5039 from radio to VHE from Dubus
\cite{dubus06}. The emission at $10^{15}$~Hz corresponds to thermal emission
from the optical companion. The rest of the emission is non-thermal emission
produced by the interaction with the compact object. 
\label{fig:ls5039sed}}
\end{figure}

The broadband emission of LS~5039 is shown in Fig.~\ref{fig:ls5039sed}. The
output is dominated by the emission above 1~MeV, allowing us to consider the
system as a gamma-ray binary. In this work we are interested in studying the
behavior of the radio emission at mas scales. 

%------------------------------------------------------------------------------
%------------------------------------------------------------------------------

\section{Possible scenarios} \label{scenarios}

The HE/VHE emission of gamma-ray binaries is basically interpreted as the
result of inverse Compton upscattering of stellar UV photons by relativistic
electrons. Two excluding scenarios have been proposed to explain the
acceleration mechanism that powers the relativistic electrons. In the first one
electrons are accelerated in the jets of a microquasar powered by accretion. In
the second one they are accelerated in the shock between the relativistic wind
of a non-accreting pulsar and the wind of the stellar companion.

Microquasars are X-ray binary systems containing a black hole or a neutron star
and a companion star, from which material is accreted by the compact object.
They display bipolar relativistic collimated plasma jets. The electrons in the
jets produce synchrotron radiation when they interact with the magnetic fields
present in the jet. In this scenario the VHE emission is produced by inverse
Compton scattering when the jet particles collide with stellar UV photons. Some
models also take into account hadronic processes, in which accelerated protons
collide with stellar wind ions. For detailed models of the broadband emission
from microquasars see \cite{bosch06}, \cite{paredes06}, and \cite{romero03}.

In the non-accreting pulsar scenario the relativistic wind of a young
millisecond pulsar collides with the stellar wind of the bright companion. The
region where the wind pressures are balanced defines the standoff distance,
which is close to the compact object. The high energy emission is produced in
this region, while a nebula of accelareted particles forms behind the pulsar.
The cooling processes of these accelarated particles along the adiabatically
expanding flow produces the non-thermal broadband emission. The morphology is
similar to the one expected in isolated pulsars moving through the ISM.
Information on this scenario can be found in \cite{maraschi81}, \cite{dubus06},
and \cite{sierpowska07}.

\section{Testing at mas scales} \label{testing}

The expected behavior of the radio emission at mas scales is different in each
scenario, allowing us to test the scenarios by means of high resolution radio
images. In the microquasar scenario we have a central core with extended
jet-like radio emission. The projection effects and the Doppler boosting of the
relativistic jets produce a flux and distance asymmetry that can be measured
\cite{mirabel99}, \cite{fender06}. The extended emission can be one sided for
highly relativistic jets or if the direction of the flow is near the line of
sight. The direction of the jets should remain constant during an orbital
cycle, although it can display long-term precession. The dense stellar wind of
the bright companion can interact with the relativistic jets, producing the
bending and disruption of the jets \cite{perucho08}.

In the non-accreting pulsar scenario the shocked material is contained by the
stellar wind behind the pulsar, producing a 'bow' shaped nebula extending away
from the stellar companion. As a consequence of this, the tail of the flow at a
few AU follows an elliptical path during the orbital cycle, while at larger
scales we see the accumulated contribution of electrons accelerated during
several orbital cycles. At mas scales, it is expected that the direction of the
extended emission changes with the pulsar's orbital motion. The peak of the
radio emission, located at a few AU behind the pulsar, should follow an
elliptic orbit $\sim$10 times bigger than the size of the orbit of the system
(the size depends on the magnetization of the pulsar wind), \cite{dubus06}.

\section{Previous radio observations} \label{previous}

The unresolved radio emission of LS~5039 is persistent, non-thermal, and
variable, although no strong radio outburst or periodic variability have been
detected so far \cite{ribo99,clark01}. A multi-wavelength study between 1.4~GHz
and 15~GHz using the VLA in its A configuration was performed in Mart\'{\i} et
al.\  \cite{marti98}, always showing an unresolved point-like source
($\leq0.1^{\prime\prime}$). The source spectral index resulted to be
$\alpha=-0.46\pm0.01$, very suggestive of non-thermal radio emission. No radio
pulses have been detected at 1.4~GHz \cite{morris02}. The system might still
contain a pulsar, but in this case free-free absorption with the stellar wind
would prevent us to detect the pulsations. New observations at higher
frequencies may be required to search for pulsations. 

\begin{figure} 
\center 
\includegraphics[angle=0,scale=0.4]{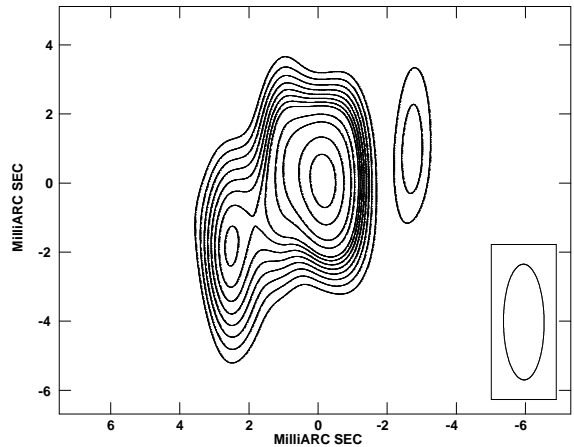}
\caption{VLBA radio map of LS~5039 at 5~GHz obtained on 1999 May 8. The
contours shown correspond to 6, 8, 10, 12, 14, 16, 18, 20, 25, 30, 40, and 50
times  0.085~mJy~beam$^{-1}$, the rms noise. The ellipse at the bottom right
corner represents the half-power beam width of the synthesized beam,
3.4~mas$\times1.2$~mas with PA of $0^{\circ}$. North is up and east is to the
left. The orbital phase of the system was in the range 0.12--0.15 during the
observations. \label{fig:bp051}}
\end{figure}

The discovery of the radio structure of the source was presented in Paredes et
al.\ \cite{paredes00}, where observations at 5~GHz were performed using the
VLBA and the VLA in its phased array mode. The observation took place on May 8,
1999, which correspond to orbital phases in the range 0.12--0.15, when computed
with the ephemeris from \cite{casares05}. The final synthesis map (see
Fig.~\ref{fig:bp051}) shows two-sided extended emission emerging from a central
core. The deconvolved angular size of the core is 2~mas, and it accounts for
the 80\% of the total flux density recovered, which is 16~mJy. The extended
emission is aligned towards a direction with Position Angle (PA)
$=125^{\circ}$, measured from north to east. The source extends over 6~mas on
the plane of the sky. Three main components can be fitted by the following
model: a central core of 12.8~mJy, a south-east component located at an angular
distance of $3.8\pm0.2$~mas from the core and with a flux density of
$2.1\pm0.1$~mJy, and a north-west component located at  $2.8\pm0.2$~mas from
the core and with a flux density of $1.0\pm0.1$~mJy. We can see in
Fig.~\ref{fig:bp051} that the core is not point-like and seems to be resolved
or to present some extended structure. However, this emission is too faint to
analyze it in the model fit.

Simultaneous observations at 5~GHz using the EVN and MERLIN were reported in
Paredes et al.\ \cite{paredes02}. The orbital phase of the system was in the
range 0.35--0.39 during the observations. The images show bipolar extended
emission emanating from a central core, like the VLBA maps. In both images the
south-east emission is brighter and larger than the north-west one. The
extended emission was detected up to 35~mas (100~AU) and 175~mas (400~AU) in
PA of 140$^\circ$ and 150$^\circ$ respectively, which suggest that the
extended emission could be bent.

%------------------------------------------------------------------------------
%------------------------------------------------------------------------------

\section{VLBA observations in June 2000} \label{observations}

We observed LS~5039 with the Very Long Baseline Array (VLBA) and the Very Large
Array (VLA), of the National Radio Astronomy Observatory (NRAO), at 5~GHz
frequency on 2000 June 3 and 8. The project codes for these Global VLBI
observations are GR021A and GR021B, respectively. The VLA data was correlated
twice: first as a connected interferometer, which provides low resolution
images to obtain detailed flux density information, and secondly in its phased
array observing mode, serving as an additional single antenna for VLBI. 

The two observing sessions, hereafter run~A and run~B, spanned from 4:30 to
12:30~UT on the corresponding dates, and were thus centered on MJD~51698.4 and
MJD~51703.4, respectively. The two runs, performed 5~days apart, took place in
consecutive orbital cycles. The orbital phases of the system were in the range
0.43--0.51 for run~A and in the range 0.71--0.79 of the following orbital cycle
for run~B. During run~A, the compact object was arriving to the apastron,
whereas during run~B it was located after the inferior conjunction.

The observations were performed switching between the target source LS~5039 and
the phase reference calibrator J1825$-$1718, with cycling times of 5.5 minutes.
The 3.5~Jy ICRF source J1911$-$2006, located at 11.9$^\circ$ from LS~5039, was
observed every 22 minutes in order to monitor the performance of the
observations. The position used hereafter for the phase reference source
J1825$-$1718, obtained by means of dedicated geodetic VLBI observations, is
$\alpha_{\rm J2000.0}=18^{\rm h} 25^{\rm m} 36\rlap.^{\rm s}53237$
($\pm3.0$~mas) and $\delta_{\rm J2000.0}=-17^\circ 18^\prime 49\rlap.{''}8534$
($\pm4.5$~mas). This position was provided by observations from the joint
NASA/USNO/NRAO geodetic/astrometric program. The coordinates are in the frame
of ICRF-Ext.1. However, this information was not available at the time of
correlation, which was performed for a calibrator position shifted by
$\Delta\alpha=+57.0$~mas and $\Delta\delta=+21.4$~mas. In the case of LS~5039,
due to its proper motion, the source was found to be $\Delta\alpha=+7.6$~mas
and $\Delta\delta=-16.2$~mas away from the correlated position. 

The post-correlation data reduction was performed using the Astronomical Image
Processing System ({\sc aips}) software package, developed and maintained by
NRAO. Corrections of $\sim$0.1~mm were applied to the positions of some of the
VLBA antennas, although the VLA position had to be corrected by 0.43~m,
according to later geodetic measurements. The positions of J1825$-$1718 and
LS~5039 were corrected using the task {\sc clcor}. As recommended for
phase-referencing experiments, we applied ionospheric and Earth Orientation
Parameters corrections to the visibility data using the task {\sc clcor}. A
priori visibility amplitude calibration was done using the antenna gains and
the system temperatures measured at each station. The fringe finder was then
used to calibrate the instrumental phase and delay offsets between the
different intermediate frequency (IF) channels. The fringe fitting ({\sc
fring}) of the residual delays and fringe rates was performed for all the
sources. Very good solutions were found for tha calibrators J1911$-$2006 and
J1825$-$1718. Fringes for 15 and 25\% of the baselines were missing for the
target source LS~5039 and for the astrometric check source J1837$-$1532,
respectively. Typical data inspection and flagging were performed. An
independent reduction of the VLA data was performed using standard procedures
within {\sc aips}.

\section{Results} \label{results}

\begin{figure}[t!]
\center
\includegraphics[angle=-90,scale=0.7]{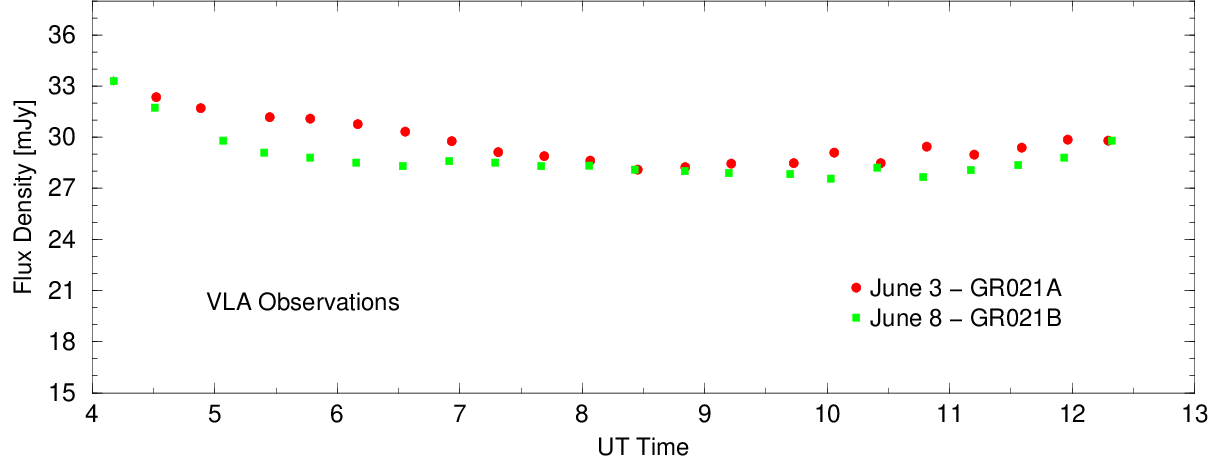}
\caption{Flux density as a function of time of the self-calibrated VLA
observations for both 2000 June runs. Red circles correspond to run~A and green
squares to run~B. The flux density errors are smaller than the symbols.
\label{fig:vla}}
\end{figure}

The VLA data correlated as an independent interferometer provided low
resolution images and very precise flux density information. The VLA data of
LS~5039 were compatible with a point-like source for the obtained synthesized
beam of $5.5^{\prime\prime}\times3.7^{\prime\prime}$ in PA $\sim-4^\circ$.
We measured the flux density of the source every 30 minutes and obtained a mean
of 29.4~mJy with a standard deviation of $\sigma$=1.1~mJy during the 8 hours of
run~A, and 28.4~mJy with $\sigma$=0.7~mJy for run~B. The obtained fluxes are
plotted in Fig.~\ref{fig:vla}.

%------------------------------------------------------------------------------
%------------------------------------------------------------------------------

\begin{figure}[t!]
\center
\includegraphics[angle=0,scale=0.7]{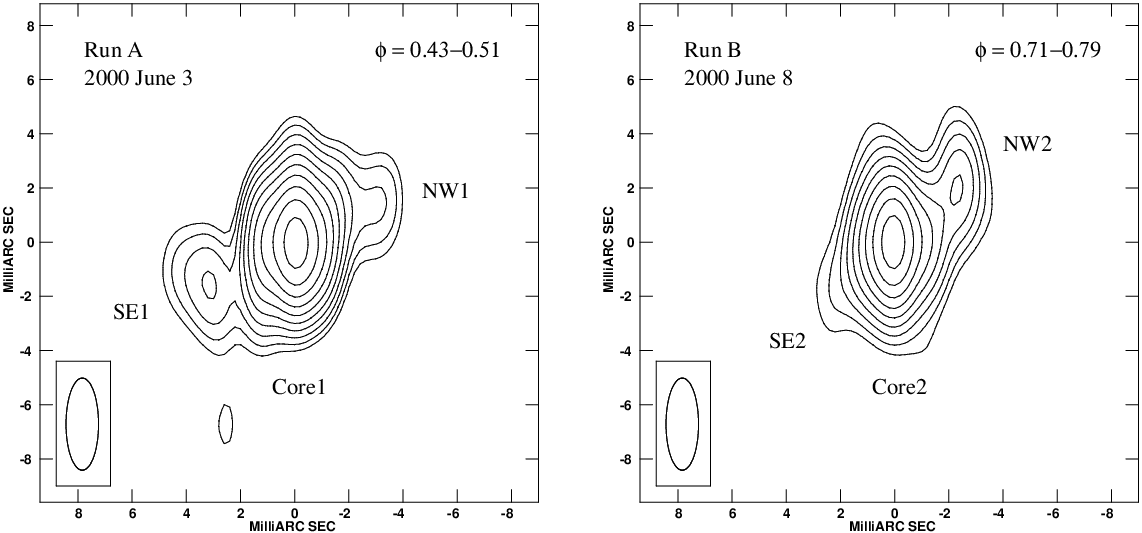}
\caption{VLBA+phased VLA self-calibrated images of LS~5039 at 5~GHz obtained on
2000 June 3 (left) and 8 (right). North is up and east is to the left. Axes
units are in mas, and the (0,0) position corresponds to the source peak in each
image. The convolving beam, plotted in the lower left corner, has a size of
3.4$\times$1.2~mas in PA of $0^{\circ}$. The first contour corresponds to 5
times the r.m.s.\ noise of the image (0.08 and 0.11~mJy~beam$^{-1}$ for run~A
and B, respectively), while consecutive ones scale with $2^{1/2}$. The dates
and orbital phases are quoted in the images. There is extended radio emission
that appears nearly symmetric for run~A and clearly asymmetric for run~B, with
a small change of $\sim10^{\circ}$ in its position angle.
\label{fig:vlba}}
\end{figure}

We show the final VLBA+phased VLA self-calibrated images in
Fig.~\ref{fig:vlba}. The image obtained for run~A displays a central core and a
bipolar and nearly symmetric extended emission with
PA~$\simeq116\pm2^{\circ}$, with brightest component towards the south-east.
The total flux density recovered by the VLBA, obtained with the task {\sc
tvstat} within {\sc aips}, is $\sim$25~mJy, representing $\sim$85\% of the VLA
one. The peak flux density of the core is 10.5~mJy~beam$^{-1}$. The image is
similar to the one obtained with the same array in 1999 May 8, corresponding to
orbital phases 0.12--0.15, which showed a slightly more asymmetric extended
emission in PA$\sim125^{\circ}$, as can be seen in Fig.~\ref{fig:bp051}. In
contrast, the image obtained for run~B displays a core and a bipolar but
clearly asymmetric structure with PA $\simeq128\pm2^{\circ}$ and the
brightest component towards the north-west. The total flux density recovered is
$\sim$24~mJy, or $\sim$85\% of the VLA one. The peak flux density of the core
is 10.5~mJy~beam$^{-1}$, as in run~A. To characterize the extended emission we
used \texttt{UVFIT} and \texttt{JMFIT} within {\sc aips}, as well as model
fitting tools within {\sc Difmap} to check the reliability of the obtained
results. The preferred model for run~A data consisted of 3 Gaussian components
to account for the core (Core1), the south-east (SE1), and the north-west (NW1)
components. A similar model could fit the data for run~B, although the
south-east (SE2) component is marginally fitted. The fitted parameters are
quoted in Table~\ref{tab:parameters}. See also \cite{ribo08}.

We splitted the 8~hours of data in each run into 4-hour data sets. For each of
them we model fitted the visibilities of the self-calibrated data. Although not
all the components were well fitted (NW1 is marginally fitted and SE2 is not
detected in the partial images), no significant morphological differences are
measured between the two halves in any of the two runs. The peak position of
the component SE1 with respect to Core1 is stable in 4 hours within the errors
($\sigma_{\alpha}^{\rm A}=0.31$~mas, $\sigma_{\delta}^{\rm A}=0.62$~mas). For
run~B the distance between Core2 and NW2 is also stable in 4 hours within the
errors ($\sigma_{\alpha}^{\rm B}=0.55$~mas, $\sigma_{\delta}^{\rm
B}=0.25$~mas).

There are no reliable astrometric results obtained with the phase-referenced
images of current data. The resolution of the images of LS~5039 and the
check-source is limited by the scatter broadening of the phase-reference
source. Furthermore, the offset of 60.9~mas in the correlated position of
J1825$-$1819 produces an instrumental drift of the position of the target
source of $2.8\pm0.2$~mas during the 8-h runs, in the same direction as the
line joining the positions of the source and the phase-reference calibrator
(see \cite{ribo08}). Accounting for these errors, the measured
position of LS~5039 for run~A is $\alpha_{\rm J2000.0}=18^{\rm h} 26^{\rm m}
15\rlap.^{\rm s}05653\pm0\rlap.^{\rm s}00001$ (or $\pm$0.15~mas), and
$\delta_{\rm J2000.0}=-14^\circ 50^\prime 54\rlap.{''}2564\pm0\rlap.{''}0015$
(or $\pm$1.5~mas). For run~B the source is splitted in two components during
the observation, preventing us to obtain an accurate position of the source.
%-----------------------------------------------------------------------------
%-----------------------------------------------------------------------------

%-----------------------------------------------------------------------------
\begin{table}[t!] %table1
\begin{center}
\caption{Parameters of the Gaussian components fitted to the data. Columns~3
and 4 list the peak and integrated flux densities of each component. Columns~5
to 8 list the polar and Cartesian coordinates of the components with respect to
the peak position. The PA is positive from north to east.}
\label{tab:parameters} 
\begin{tabular}{l@{~~}l@{~}r@{ $\pm$ }l@{~~}r@{ $\pm$ }l@{~~}r@{ $\pm$ }l@{~~}r@{ $\pm$ }l@{~~}r@{ $\pm$ }l@{~~}r@{ $\pm$ }l}
\hline
\hline
Run & Comp. & \multicolumn{2}{l}{Peak $S_{\rm 5~GHz}$} & \multicolumn{2}{c}{$S_{\rm 5~GHz}$} & \multicolumn{2}{c}{$r$}   & \multicolumn{2}{c}{PA}         & \multicolumn{2}{c}{$\Delta\alpha$} & \multicolumn{2}{c}{$\Delta\delta$}  \\
    &       & \multicolumn{2}{l}{[mJy~beam$^{-1}$]}    & \multicolumn{2}{c}{[mJy]}           & \multicolumn{2}{c}{[mas]} & \multicolumn{2}{c}{[$^{\circ}$]} & \multicolumn{2}{c}{[mas]}          & \multicolumn{2}{c}{[mas]}           \\
\hline
%\multicolumn{19}{c}{Run~A} \\
%\hline
A   & Core1 & 10.54 & 0.08 & 20.0 & 0.2 & \multicolumn{2}{c}{---} & \multicolumn{2}{c}{---} & \multicolumn{2}{c}{---} & \multicolumn{2}{c}{---}  \\
    & SE1   &  1.11 & 0.08 &  2.6 & 0.2 & 3.67  & 0.08 & 115.9   & 1.7 &    3.30 & 0.07 &  $-$1.60 & 0.12  \\
    & NW1   &  0.88 & 0.08 &  1.5 & 0.2 & 3.29  & 0.09 & $-$63   & 2   & $-$2.92 & 0.08 &     1.52 & 0.14  \\
\hline
%\multicolumn{19}{c}{Run~B} \\
%\hline
B   & Core2 & 10.45 & 0.11 & 17.6 & 0.3 & \multicolumn{2}{c}{---} &\multicolumn{2}{c}{---} &\multicolumn{2}{c}{---} &\multicolumn{2}{c}{---}  \\
    & SE2   &  0.75 & 0.11 &  1.8 & 0.4 & 2.8~~ & 0.2  &     129 & 5   &    2.17 & 0.13 & $-$1.8~~ & 0.3   \\
    & NW2   &  2.22 & 0.11 &  3.9 & 0.3 & 2.94  & 0.06 & $-$52.2 & 1.4 & $-$2.32 & 0.04 &     1.80 & 0.09  \\
\hline
\end{tabular}
\end{center}
\end{table}
%-----------------------------------------------------------------------------

\section{Discussion} \label{discussion}

The observations of LS~5039 reported here, obtained with the VLBA on two runs
separated by 5 days, show a changing morphology at mas scales. In both runs
there is a core component with a constant flux density within errors, and
elongated emission with a PA that changes by $12\pm3^{\circ}$ between both
runs. The brightest component in run~A is towards south-east, and in run~B
towards north-west. The source is nearly symmetric in run~A and asymmetric in
run~B (see Fig.~\ref{fig:vlba}).

In the microquasar scenario, and assuming ballistic motions of adiabatically
expanding plasma clouds without shocks \cite{mirabel99}, the morphology of
run~A can be interpreted as a double-sided jet emanating from a central core
with the southeast component as the approaching one, whereas in run~B the
northwest component would be the approaching one. We can compute the projected
bulk velocity of the jets from the distance and flux asymmetry. The results,
shown in Table~\ref{tab:jets}, are compatible with a mildly relativistic jet
with a projected bulk velocity of 0.1c. However, the distances from Core2 to
the components NW2 and SE2 are very similar and do not imply any significant
relativistic motion.

We can compute, for different $\theta$ angles, the expected displacement of the
approaching components with respect to the respective cores in 4 hours using
the measured value for the flux asymmetry. The lack of proper motions (see
upper limits in Sect.~\ref{results}) implies that, for the measured flux
asymmetries, the SE1 jet should be pointing at $\theta<48^{\circ}$, and the NW2
jet at $\theta<45^{\circ}$. In this context, jet precession is needed to
explain this behavior. If the precession axis is close to the plane of the sky,
as in SS~433 \cite{blundell04}, the precession cone should have a semi-opening
angle $>45^{\circ}$ to fulfill the $\theta$ limits quoted above. If the
precession axis is close to the line of sight, a small precession of a few
degrees could explain the images of runs A and B. However, in both cases the
PA of the jet should vary considerably, in contrast to the small range
covered by all observed values at mas scale, between 115 and 140$^{\circ}$
\cite{paredes00,paredes02}.

%------------------------------------------------------------------------------
\begin{table}
%\begin{center}
\caption[]{Projected bulk velocities from the distance asymmetry of the
components, and the flux asymmetry in the case of a persistent jet and a
discrete ejection.}
\begin{center}
\begin{tabular}{cccccccccc}
\noalign{\smallskip} \hline \hline \noalign{\smallskip}
               & &  \multicolumn{2}{c}{Distance asymmetry} & &  \multicolumn{5}{c}{Flux asymmetry} \\
\noalign{\smallskip} \cline{3-4} \cline{6-10} \noalign{\smallskip}      
              & &                      &                   & & \multicolumn{2}{c}{continuous (k=2)}    & &  \multicolumn{2}{c}{discrete (k=3)}     \\
 run        & &  $\beta\cos\theta$   &  $\theta~(^{\circ})$ & & $\beta\cos\theta$  &  $\theta~(^{\circ})$ & &  $\beta\cos\theta$  &  $\theta~(^{\circ})$ \\
\noalign{\smallskip} \hline \noalign{\smallskip}        
 A         &  &  $0.06\pm0.02$      &  $<87$            & & $0.11\pm0.03$      &  $<84$            & &  $0.08\pm0.02$      &  $<85$       \\
 B         &  &  \multicolumn{2}{c}{symmetric}          & & $0.15\pm0.05$      &  $<81$            & &  $0.11\pm0.03$      &  $<84$       \\
\noalign{\smallskip} \hline
\end{tabular}
\end{center}
\label{tab:jets}
\end{table}
%------------------------------------------------------------------------------

Alternatively, the morphology detected in run~B could be the result of a
discrete ejection where Core2 is the approaching component and NW2 the receding
one, while there is no radio emission at the origin of the ejection. However,
large X-ray and radio flux density variations are expected during discrete
ejections (see \cite{fender06} and references therein), while the peak and
total radio flux densities of LS~5039 are strikingly constant (see also
\cite{ribo99} and \cite{clark01}) and there is no evidence of an X-ray flare in
11.5 years of {\it RXTE}/ASM data. Although not yet explored in detail, the
morphology changes can be due to the interaction between the jets and a clumpy
or dense stellar wind \cite{perucho08}.

In the non-accreting pulsar scenario, the different morphologies we have
detected at different orbital phases could be due to the change of the relative
positions between the pulsar and the companion star along the orbit (see
details in \cite{dubus06}). Observations at different orbital phases have
always revealed a very similar PA for the extended emission, which would
correspond to an inclination of the orbit as seen nearly edge on (90$^\circ$).
On the other hand, the absence of X-ray eclipses places an upper limit of
$i\lesssim 75^{\circ}$ in this scenario. Therefore, these two restrictions
imply an inclination angle that should be close to the upper limit of
75$^{\circ}$. 

In conclusion, a simple and shockless microquasar scenario cannot easily
explain the observed changes in morphology. On the other hand, an
interpretation within the young non-accreting pulsar scenario requires the
inclination of the binary system to be very close to the upper limit imposed by
the absence of X-ray eclipses. Precise phase-referenced VLBI observations
covering a whole orbital cycle are necessary to trace possible periodic
displacements of the peak position, expected in this last scenario, and to
obtain morphological information along the orbit. These will ultimately reveal
the nature of the powering source in this gamma-ray binary.

\acknowledgments{
The NRAO is a facility of the National Science Foundation operated under
cooperative agreement by Associated Universities, Inc. J.M., M.R., J.M.P., and
J.M. acknowledge support by DGI of the Spanish Ministerio de Educaci\'on y
Ciencia (MEC) under grants AYA2007-68034-C03-01 and AYA2007-68034-C03-02 and
FEDER funds. J.M. was supported by MEC under fellowship BES-2008-004564. M.R.
acknowledges financial support from MEC and European Social Funds through a
\emph{Ram\'on y Cajal} fellowship. J.M. is also supported by Plan Andaluz de
Investigaci\'on of Junta de Andaluc\'{\i}a as research group FQM322.
}

\end{document}